\documentclass[aps,prl,reprint,groupedaddress]{revtex4-1}
\usepackage[dvipdfmx]{graphicx}
\usepackage{url}

\usepackage{amsmath}
\usepackage{booktabs}
\usepackage{tabularx}
\title{}
\begin{document}

\title{Predictions of  tertiary stuctures of $\alpha$-helical membrane proteins by replica-exchange method with consideration of helix deformations}

\author{ Ryo Urano} \affiliation{Department of Physics, Graduate School of Science, Nagoya University, Nagoya, Aichi 464-8602}  \author{Hironori Kokubo} \email{Present address: Takeda Pharmaceutical, Fujisawa, Kanagawa 251-8585} \affiliation{ Depertment of Functional Molecular Science, The Graduate University for Advanced Studies, Okazaki, Aichi  444-8585} \author{Yuko Okamoto} \email[]{okamoto@phys.nagoya-u.ac.jp} \affiliation{Department of Physics, Graduate School of Science, Nagoya University, Nagoya, Aichi 464-8602} \affiliation{Center for Computational Science, Graduate School of Engineering, Nagoya University, Nagoya, Aichi 464-8603}\affiliation{Structural Biology Research Center, Graduate School of Science, Nagoya University, Nagoya, Aichi 464-8602}\affiliation{Information Technology Center, Nagoya University, Nagoya, Aichi 464-8601}

\begin{abstract}
We propose an improved prediction method of the tertiary structures of $\alpha$-helical membrane proteins based on the replica-exchange method by taking into account helix deformations. Our method allows wide applications because transmembrane helices of native membrane proteins are often distorted. In order to test the effectiveness of the present method, we applied it to the structure predictions of glycophorin A and phospholamban. The results were in accord with experiments. 
\end{abstract}


\maketitle


\section*{Introduction}
\label{sec-1}
Membrane proteins are  fundamental for life, and their structures and dynamics are essential for their biological functions. About 30 \% of proteins encoded in genomes are estimated to be of membrane proteins by bioinformatics.\cite{krogh2001predicting,sawada2012biological}
Although membrane-protein folding has been studied extensively by experiments,\cite{fiedler_protein_2010,von_heijne_introduction_2011} only about 2 \% in whole known structures in PDB are membrane proteins, because biomembrane environment makes crystallization very difficult.\cite{pdbtm1,pdbtm2,mpdb,opm,pdbmpks} 
Thus, simulation studies are getting more important (for previous simulations, see, for instance, \cite{taylor1994method,suwa1995continuum,adams1996improved,pappu1999potential,hirokawa2000triangle,vaidehi2002prediction,bu2007membrane,miyashita2009transmembrane,leguebe2012hybrid}).

However, simulations often suffer from sampling insufficiency and the efficient sampling methods like generalized-ensemble algorithms and/or the reduction of the size of systems are required. In particular, replica-exchange method (REM) \cite{rem1,swendsen1986replica,rem3,rem4} and its extensions are often used in generalized ensemble algorithms due to their efficiency, parallelization ease, and  usability (for reviews,  see, e.g., Refs.\cite{mitsutake_generalizedensemble_2001,rem_iba2001extended}).  

One of useful approaches to reduce system sizes is to employ an implicit membrane model, which mimics some elements of membrane properties such as dielectric profile, chain order, pressure profile, and intrinsic curvature by parameters for electrostatic solvent free energy.\cite{eef1lazaridis_effective_1999,gbim_implicit_2003,hdgb1tanizaki_generalized_2005}
While these methods are mainly based on the free energy difference between solvent and solute, simpler implicit membrane model was introduced previously, where transmembrane helices keep a helix structure and are always restricted within membrane regions during folding, which greatly reduces the effort for the search in the conformational space during folding processes.\cite{Kokubo2004397}
This model assumed that the native structure of  membrane proteins can be predicted by helix-helix interactions between transmembrane helices with fixed helix structures, and that the membrane environment constraints the regions where helices can exist (namely, within membranes)
and stabilizes transmembrane helix structures. 
This model is supported by many experimental data such as those leading to the two-stage model, in which each helix structure is formed first, and they aggregate each other by helix-helix packing in membrane protein folding to reach the native conformation (for a review, see Ref. \cite{popot_h_2000}).

The previous method  \cite{Kokubo2004397,kokubo_prediction_2004,kokubo_classification_2004,Kokubo2004168,kokubo_analysis_2009} could predict the native structures by the REM simulation using known native helix structures (for a review, see Ref. \cite{kokubo_replica-exchange_2006}).
However, if the native structures consist of distorted helix structures, the previous prediction method will not work because the method treated helix structures as rigid bodies. It is actually known from experimental structures in PDB that transmembrane helices are distorted or bent in about 25 \% of all transmembrane helix structures. \cite{hall2009position}

Therefore, in this article, we propose a new treatment of helix structures by taking into account helix distortions and kinks instead of treating them as rigid bodies. We tested our new prediction method for native structures. Our test systems consist of the case with only ideal helix structures and that with a distorted helix structure.

This article is organized as follows. In Section 2, we explain the details of our methods. The potential energy function used for our new models and the method to introduce the helix kinks are described. In Section 3, we show the results of the REM simulation applied to glycophorin A and phospholamban. After we check that REM simulation are properly performed, the free energy minimum states are identified by the  principal component analysis. Finally, Section 4 is devoted to the conclusions. 

\section*{Methods}
\label{sec-2}
\subsection*{Simulation details}
\label{sec-2-1}
We first review our previous method\cite{Kokubo2004397,kokubo_prediction_2004,kokubo_classification_2004,Kokubo2004168,kokubo_analysis_2009}.
Only the transmembrane helices are used in our simulations, and loop regions of membrane proteins as well as lipid and water molecules are neglected. Our assumptions are that a role of water is to push the hydrophobic transmembrane regions of membrane proteins into the lipid bilayer and that a major role of lipid molecules is to prepare a hydrophobic environment and construct helix structures in the transmembrane regions. Loop regions of membrane proteins are often outside the membrane and we assume that they do not directly affect the structure of transmembrane regions. 
Due to the difference in surface shapes of helices and lipids, the stabilization energy for helix-helix packing will be larger than that for helix-lipid packing. Therefore, water, lipids, and loop-region of proteins are not treated explicitly in our simulations, although the features of membrane boundaries are taken into account by the constraint conditions below.

We update configurations with a rigid translation and rotation of each $\alpha$-helix and torsion rotation of side-chains by Monte Carlo (MC) simulations. We use MC method although we can also use molecular dynamics in principle. There are 2$N_{\rm H}$ + $N_{\rm SD}$ kinds of MC move sets, where $N_{\rm H}$ is the total number of transmembrane helices in the protein, and $N_{\rm SD}$ is the total number of dihedral angles in the side-chain of $N_{\rm H}$ helices.

We add the following three elementary harmonic constraints to the original potential energy function. The constraint function is given by

\begin{eqnarray}
E_{\rm constr} = E_{\rm constr1} + E_{\rm constr2} + E_{\rm constr3},
\label{const-ene}
\end{eqnarray}
where each term on the right-hand side is defined as follows: 
\begin{eqnarray}
E_{\rm constr1} = \sum_{i=1}^{N_{\rm H}-1} k_1~ \theta \left( r_{i,i+1}-d_{i,i+1} \right)
\left[ r_{i,i+1}-d_{i,i+1} \right]^2,
\label{const-ene1}
\end{eqnarray}

\begin{eqnarray}
E_{\rm constr2} &= \displaystyle{\sum_{i=1}^{N_{\rm H}}} \left\{ k_2~ \theta \left( \left| z^{\rm L}_{i}-z^{\rm L}_{0} 
\right| -d^{\rm L} \right)
\left[ \left| z^{\rm L}_{i}-z^{\rm L}_{0} \right| -d^{\rm L}\right]^2\right. \nonumber \\
&+\left.k_2~ \theta \left( \left| z^{\rm U}_{i}-z^{\rm U}_{0} \right| -d^{\rm U} \right)
\left[ \left| z^{\rm U}_{i}-z^{\rm U}_{0} \right| -d^{\rm U} \right]^2 \right\}, 
\label{const-ene2}
\end{eqnarray}
\begin{eqnarray}
E_{\rm constr3} = \sum_{{\rm C}_{\alpha}} k_3~ \theta \left( r_{{\rm C}_{\alpha}}-d_{{\rm C}_{\alpha}} \right) 
\left[ r_{{\rm C}_{\alpha}}-d_{{\rm C}_{\alpha}} \right]^2.
\label{const-ene3}
\end{eqnarray}

$E_{\rm constr1}$ is the energy that constrains pairs of 
adjacent helices along the amino-acid chain not to be apart from each other 
too much (loop constraints).
$r_{i,i+1}$ is the distance between the C atom of the C-terminus of the $i$-th 
helix and the C$^{\alpha}$ atom of the N-terminus of the $(i+1)$-th helix, and 
$k_1$ and $d_{i,i+1}$ are the force constant and the central value constant 
of the harmonic constraints, respectively, and
$\theta(x)$ is the step function:
\begin{eqnarray}
\theta(x)=\left\{ \begin{array}{ll}
1~, & {\rm for} ~x \geq 0~, \\
0~, & {\rm otherwise}~. \\
\end{array} \right.
\label{step-func}
\end{eqnarray} 
This term has a non-zero value only when the 
distance $r_{i,i+1}$ becomes longer than $d_{i,i+1}$. 
Only the structures in which the distance between neighboring helices in the amino-acid sequence is 
short are searched 
because of this constraint term.

$E_{\rm constr2}$ is the energy that constrains 
helix N-terminus and C-terminus to be located near membrane boundary planes. 
Here, the z-axis is defined to be the direction perpendicular to the membrane boundary planes.
$k_2$ is the force constant of the harmonic constraints.
$z^{\rm L}_{i}$ and 
$z^{\rm U}_{i}$ are the z-coordinate values of 
the C$^{\alpha}$ atom of the 
N-terminus or C-terminus of the $i$-th helix near the fixed lower 
membrane boundary and the upper membrane boundary, respectively.
$z^{\rm L}_0$ and $z^{\rm U}_0$ are the fixed lower boundary
z-coordinate value and the upper boundary z-coordinate value
of the membrane planes, respectively.
$d^{\rm L}$ and 
$d^{\rm U}$ are the corresponding central value constants
of the harmonic constraints.
This term has a non-zero value only when 
the C$^{\alpha}$ atom of the 
N-terminus or C-terminus of the $i$-th helix
are apart more than $d^{\rm L}_{i}$ (or $d^{\rm U}_{i}$). 
This constraint energy was introduced so that the helix ends are not too 
much apart from the membrane boundary planes.

$E_{constr3}$ is the energy that 
constrains all C$^{\alpha}$ atoms within the sphere (centered at the origin) 
of radius $d_{{\rm C}_{\alpha}}$.
 $r_{{\rm C}_{\alpha}}$ is the 
distance of C$^{\alpha}$ atoms from the origin, and $k_3$ and
 $d_{{\rm C}_{\alpha}}$ are the force constant and the central value constant
of the harmonic constraints, respectively.
This term has a non-zero value only when C$^{\alpha}$ atoms go out of 
this sphere and is introduced so that the center of mass of the 
molecule stays near the origin. 
The radius of the sphere $d_{{\rm C}_{\alpha}}$ is set 
to a large value in order to 
guarantee that a wide conformational space is sampled.

 These constraints are considered to be a simple implicit membrane model which mimics membrane environment during membrane protein folding.
 Moreover, all constraints limit the conformational space of proteins to improve sampling and are useful when we use limited computational resources. 
In summary, this procedure is consistent with the two-stage model, and it assumes that  side-chain flexibility is essential in their folding. 
Because backbone structures of main  chain are treated as rigid bodies in the previous method, the method can not be applied if transmembrane helices are distorted.


However, most of transmembrane helix structures in PDB have distorted or bent helix structures. We, therefore, need to treat the deformations of backbone helix structures during simulations.  
Namely,  the $\phi$ and $\psi$ torsion rotations and concerted rotation of backbone are used to reproduce the distorted helix structures of experimental structures from the initial ideal helix structures in Monte Carlo move sets. Here, we also update configurations with a rotation of torsion angles of backbones by directional manipulation and concerted rotation.\cite{dinner_local_2000,go_ring_1970,dodd_concerted_1993,wedemeyer1999exact,coutsias2004kinematic} 
There are 2$N_{\rm H}$ + $N_{\rm SD}$ +$N_{\rm BD}$ +$N_{\rm CR}$ kinds of MC moves now, where $N_{\rm BD}$ is the total number of $(\phi, \psi)$ torsion angles in the helix backbones, and $N_{\rm CR}$ is the total number of the combination of seven successive backbone torsion angle by the concerted rotation in the helix backbone.
One MC step in this article is defined to be an update of one of these degrees of freedom, which is accepted or rejected according to the Metropolis criterion.

In order to keep helix conformations of the distortions, we introduce the fourth constraint term as follows:
\begin{eqnarray}
E_{\rm constr} = E_{\rm constr1} + E_{\rm constr2} + E_{\rm constr3} + E_{\rm constr4}, 
\label{const-ene_new}
\end{eqnarray}
\begin{equation}
\begin{split}
E_{\rm constr4} &= \sum_{j=1}^{ N_{BD}} k_4   \theta(\mid \phi_j^{} - \phi_0 \mid
    -\alpha_j^{\phi}) (\mid \phi_j^{} - \phi_0\mid -\alpha_j^{\phi})^2 \\
 &  +\sum_{j=1}^{ N_{BD}} k_5  \theta(\mid \psi_j^{} - \psi_0 \mid
    -\alpha_j^{\psi}) (\mid \psi_j^{} - \psi_0\mid -\alpha_j^{\psi})^2, 
\label{const-ene4}
\end{split}
\end{equation}
where $E_{\rm constr4}$ is the newly-introduced energy term which constrains dihedral angles of main chains  within bending or kinked helix structures from ideal helix structures and prevent them from bending and distortions too much.
$\phi_j^{}$ and $\psi_j^{}$ are the main-chain torsion angles of the $j$-th residue. 
$\phi_0$ and $\psi_0$ are the fixed reference values of the harmonic constraint, $k_4$ and $k_5$ are the force constants, and 
$\alpha_j^{\phi}, \alpha_j^{\psi}$ are the central values of the harmonic constraint.

We now explain the replica-exchange method briefly.
This method prepares $M$ non-interacting replicas at $M$ different
 temperatures. While conventional canonical MC simulation is performed for each replica, temperature exchange between pairs of replicas corresponding to temperatures is attempted at a fixed interval based on the following Metropolis criterion.
Let the label $i$ (=1, $\cdots$, $M$) correspond to the replica index
 and label $m$  (=1, $\cdots$, $M$) to the temperature index. 
We represent the state of the entire system of $M$ replicas by $X = \left\{x_{m(1)}^{[1]}  , \cdots, x_{m(M)}^{[M]} \right\}$, where $x_m^{[i]} =\left\{q^{[i]}\right\}$ are the set of coordinates of replica $i$ (at temperature $T_m$), and $m=m(i)$ is the permutation of $i$.
The Boltzmann-like probability distribution for state $X$ is given by 
\begin{equation}
W_{{\rm REM}}(X)=\prod_{i=1}^M \exp{[-\beta_{m(i)} E(q^{[i]})]}.
\end{equation}
We consider exchanging a pair of temperatures $T_m$ and $T_n$, 
 corresponding to  replicas $i$ and $j$: 
\begin{eqnarray}
 X =  &\left\{ \cdots, x_m^{[i]}  , \cdots,
 x_n^{[j]}, \cdots \right\} \rightarrow  \nonumber \\
 & X^\prime =  \left\{ \cdots, x_m^{[j]}  , \cdots,
 x_n^{[i]}, \cdots \right\} .
\end{eqnarray}
The transition probability $\omega (X\rightarrow X^\prime)$ of Metropolis criterion is given by
\begin{eqnarray}
 \omega (X\rightarrow X^\prime ) &\equiv \omega (x_m^{[i]} \mid x_n^{[j]}) \nonumber \\
&= {\rm min}\left(1, \frac{W_{{\rm REM}} (X^\prime)}{W_{{\rm REM}} (X)}\right) \nonumber \\
&= {\rm min}(1,\exp(- \Delta  )) ,
\end{eqnarray}
 where $\Delta   = (\beta _m - \beta_n ) (E(q^{[j]}) - E(q^{[i]})   )$. 
Because each replica reaches various temperatures followed by replica exchange, the REM method performs a
 random-walk in temperature space during the simulation.

Expectation values of physical quantities are given as functions of temperatures by solving the multiple-histogram reweighting equations.\cite{ferrenberg_optimized_1989,kumar_weighted_1992}
The density of states $n(E)$ and dimensionless Helmholtz free energy are obtained by solving the following equations iteratively:
\begin{equation}
n(E) = { \frac{\sum\limits_{m=1}^{M}N_m(E)}{\sum\limits_{m=1}^{M}n_m e^{f_m-\beta _m E}}},
\end{equation}
and
\begin{equation}
 e^{-f_m}= \sum_{E}n(E) e^{-\beta _m E},
\end{equation}
where $N_m (E)$ and $n_m$ be the energy histogram and the total number
of samples obtained of temperature $T_m$, respectively.
After we obtained $f_m$ at each temperature, 
the expectation value of a physical quantity $A$ at any temperature $T$ is given by \cite{wham3}
\begin{equation}
 <A>_T = \frac{\sum\limits_{m=1}^{M}   \sum\limits_{x_m} A(x_m)
  \frac{1}{\sum\limits_{l=1}^{M}  n_l \exp{(f_l - \beta _l
  E(x_m))}  }  {\rm e}^{-\beta E(x_m)}  }
{ \sum\limits_{m=1}^{M}   \sum\limits_{x_m} 
  \frac{1}{\sum\limits_{l=1}^{M}  n_l \exp{(f_l - \beta _l
  E(x_m))}  }  {\rm e}^{-\beta E(x_m)}  },
\label{wham}
\end{equation}
where $x_m$ are the set of coordinates at temperature $T_m$ obtained from the trajectories of the simulation.

We analyze the simulation data by the principal component analysis (PCA).\cite{pca1,pca2,pca3,pca4,pca5,pca6} 
The structures are superimposed on an arbitrary reference structure, for example, the native structure from PDB.
The variance-covariance matrix is defined by
\begin{equation}
C_{ij} = <(q_i - <q_i>) (q_j - <q_j>)>, 
 \end{equation}
where $q_i = (q_1, q_2, q_3,\cdots , q_{3n-1}, q_{3n})=(x_1, y_1, z_1, \cdots , x_n, y_n, z_n)$
and $<\vec{q} >=\sum_{k=1}^{n} \vec{q} (k)  /n$. 
$x_i , y_i , z_i$ are Cartesian coordinates of the $i$-th atom, and $n$ is the total number of atoms.
This symmetric 3$n$ $\times$ 3$n$  matrix is diagonalized, and the eigenvectors and eigenvalues are obtained. For this calculation, we used the R program package.\cite{Rpackage2013,ihak:gent:1996,rgl}
The first superposition is performed to remove large eigenvalues
from the translations and rotations of the system, because we want to analyze
the internal differences of structures.
Therefore, this manipulation results in the smallest value close to zero for the six eigenvalues corresponding to translations and rotations of the center of geometry. The eigenvalues are ordered in the decreasing order of magnitude.
Thus, the first, second, and third principal component axes are defined as the eigenvectors corresponding to the largest, second largest, and third largest eigenvalues, respectively.
The $i$-th principal component of each sampled structure $\vec{q}$ is defined by the following inner product:
\begin{equation}
\mu _i = \nu _i \cdot (\vec{q} - <\vec{q} >) , \ \ \ \ (i=1, 2, \cdots, n), 
  \end{equation}
where $\nu _i $ is the (normalized) $i$-th eigenvector.

\subsection*{Simulation conditions}
\label{sec-2-2}
The MC program is based on CHARMM macromolecular mechanics program,\cite{brooks_charmm:_1983,hu_monte_2006} and replica-exchange Monte Carlo method was implemented in it.
In this work, we studied two membrane proteins: glycophorin A and phospholamban. Both proteins are registered in Orientation of Proteins in Membrane (OPM).\cite{opm}
The former has a dimer of an almost ideal helix structure in PDB (PDB code: 1AFO). The number of amino-acid residues in the helix is 18, and the sequence is identical  and TLIIFGVMAGVIGTILLI. 
The other has a single transmembrane helix structure in PDB (PDB code: 1FJK).
The number of amino-acid residues in the helix is 25, and the sequence is LQNLFINFCLILIFLLLICIIVMLL. The N-terminus and the C-terminus of each helix were blocked with the acetyl group and the N-methyl group, respectively. 
In the previous works, a 13-replica REM MC simulation of glycophorin A was performed with 13 replicas with the following temperatures: 200, 239, 286, 342, 404, 489, 585, 700, 853, 1041, 1270, 1548, and 1888 K. \cite{kokubo_prediction_2004,kokubo_classification_2004,kokubo_replica-exchange_2006} Although this simulation predicted the structures close to the native one successfully, the backbones structures were fixed to the ideal helix structures. In the present simulation, the flexibility of backbone helix structures is newly taken into account, and 16 replicas were used with the following temperatures: 300, 333, 371, 413, 460, 512, 571, 635, 707, 787, 877, 976, 1087, 1210, 1347, and 1499 K. The total number of MC steps was  60,000,000.
For phospholamban, 16 replicas were also used with the following temperatures: 300, 340, 386, 438, 497, 564, 640, 727, 825, 936, 1062, 1205, 1368, 1553, 1762, and 2000 K. 
The total number of MC steps was  100,000,000.
The above temperatures were chosen so that  all acceptance ratios of replica exchange are almost uniform and sufficiently large for computational efficiency.
The highest temperature was chosen sufficiency high so that no trapping in local-minimum-energy states occurs in both simulations.
Replica exchange was attempted once at every 1000 MC steps for glycophorin A and 100 MC steps for phospholamban, respectively.

We used the CHARMM19 parameter set (polar hydrogen model) for the original potential energy of the system.\cite{param19reiher1985theoretical,param19neria_simulation_1996} No cutoff was introduced to the non-bonded terms.
Each structure was first minimized subjected to harmonic restraint on all the heavy atoms.
The value of the dielectric constant was set as $\epsilon$ = 1.0, as in the previous works.\cite{kokubo_analysis_2009,Kokubo2004397,kokubo_prediction_2004,kokubo_classification_2004,Kokubo2004168}
because previous studies showed that this value was better for the predictions of transmembrane helix structures than that of $\epsilon$ = 4.0, although $\epsilon$ = 4.0 is close to the lipid environment of electrostatic potential effects. This may be due to the fact that few lipid molecules lie between helices in native transmembrane structures.
For concerted rotation we selected the backbone atoms except for those in cysteine residues. We selected 6 or 7 continuous bonds from the first atom along backbone for the driver torsion. Third bond and fifth bond were allowed to rotate following the driver bonds. 
The number of degrees of freedom in total was equal to 190 in glycophorin A and 132 in phospholamban.
We set $N_{\rm H}=2$ for glycophorin A and $N_{\rm H}=1$ for phospholamban $k_1 = 5.0$ (kcal/mol)/\AA$^2$, $d_{i,i+1} = 30.0$ \AA~ , $k_2 = 5.0$ (kcal/mol)/\AA$^2$, $k_3 = 0.5$ (kcal/mol)/\AA$^2$, $d_{{\rm C}_{\alpha}} = 50$ \AA~, $k_4=k_5=1.0$ (kcal/mol)/degrees$^2$, $\phi_0=-62$ degrees, $\psi_0=-47$ degrees, $\alpha_j^{\phi}=15$ degrees, and $\alpha_j^{\psi}=18$ degrees for our simulations.
For membrane thickness parameters, we set $z^{\rm L}_0 = -11$ \AA, $z^{\rm U}_0 = 11$ \AA, and $d^{\rm U} = d^{\rm L} = 1.0$ \AA~ for glycophorin A, and $z^{\rm L}_0 = -15$ \AA, $z^{\rm U}_0 = 15$ \AA, and $d^{\rm U} = d^{\rm L} = 1.0$ \AA~ for phospholamban.

For PCA analyses,  60,000 and 100,000  conformational data were chosen in a fixed interval at each temperature from the REM simulation for glycophorin A and phospholamban, respectively. 
We used the PDB structures (PDB codes: 1AFO for glycophorin A and 1FJK for phospholamban) as the reference structures to judge the prediction ability.

\section*{Results}
\label{sec-3}
\subsection*{Glycophorin A}
\label{sec-3-1}
\subsubsection*{Time series of various quantities}
\label{sec-3-1-1}
\begin{figure}[htb]
\centering
\includegraphics[width=.9\linewidth]{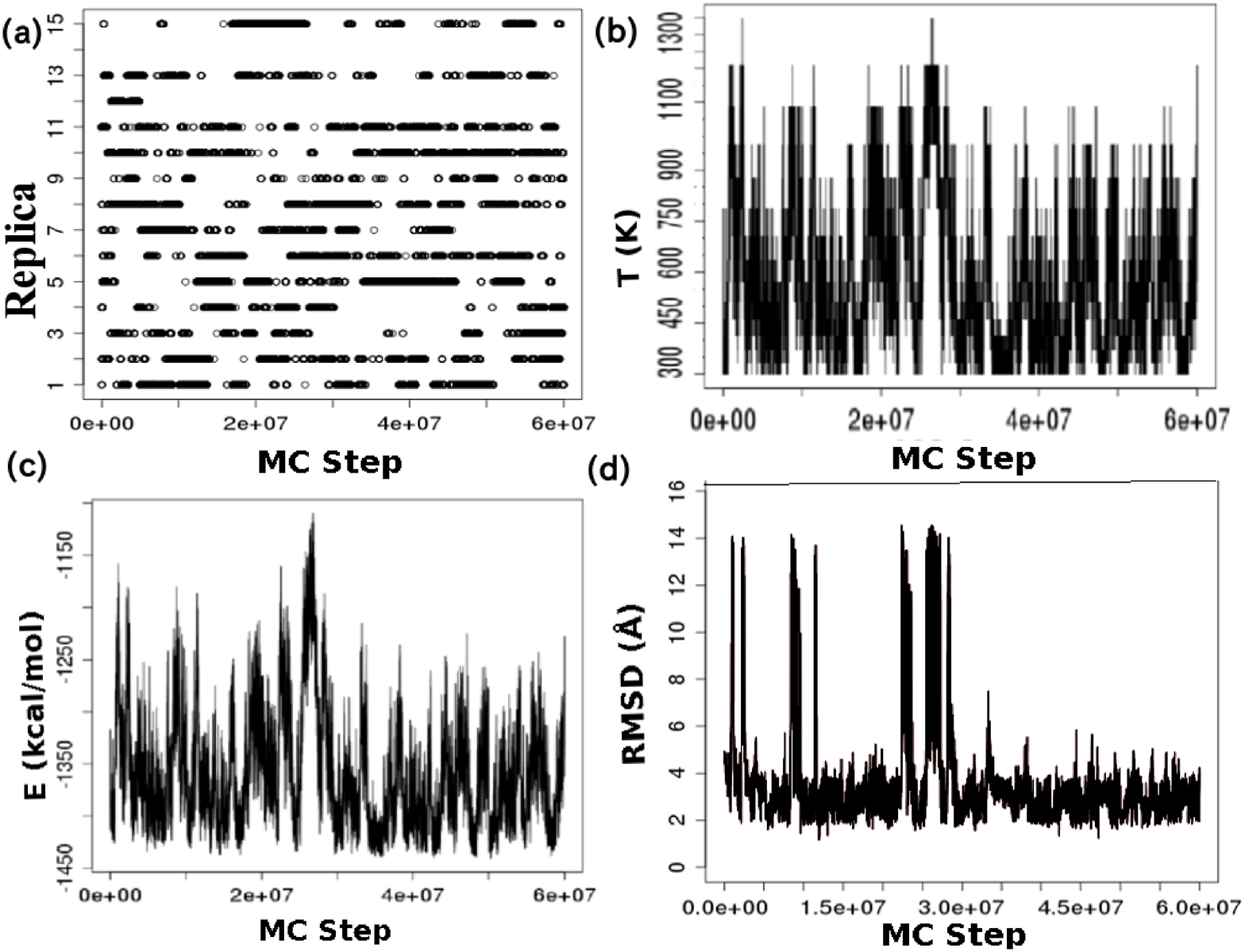}
\caption{\label{1afo-integrate}Time series of various quantities for the REM simulation of glycophorin A. (a) Time series of replica index at temperature 300 K. (b) Time series of temperature change for Replica 11. (c) Time series of total potential energy change for Replica 11. (d) Time series of the RMS deviation (in \AA{}) of all the C$^{\alpha}$ from the PDB structures for Replica 11.}
\end{figure}

We first examine how the replica-exchange simulation performed. 
Fig. \ref{1afo-integrate}(a) shows the time series of the replica index at the lowest
temperature of 300 K.
We see that the minimum temperature visited different replicas many times
during the REM simulation, and we observe a random walk in the replica space.
The complementary picture is the temperature exchange for each replica.  Fig. \ref{1afo-integrate}(b) shows the time series of temperatures for one of the replicas (Replica 11). We see that Replica 11 visited various temperatures  during the REM simulation. 
We observe random walks in the temperature space between the lowest and highest temperatures. Other replicas behaved similarly.
Fig. \ref{1afo-integrate}(c) shows the corresponding time series of  the total potential energy for Replica 11. We see a strong correlation between time series of temperatures (Fig. \ref{1afo-integrate}(b)) and that of potential energy (Fig. \ref{1afo-integrate}(c)), as is expected. We next examine how widely the conformational space was sampled during the REM simulation.
We plot the time series of the root mean-square deviation (RMSD) of all the C$^{\alpha}$ atoms from the experimental structure (PDB code: 1AFO) for Replica 11 in Fig. \ref{1afo-integrate}(d).
When the temperature becomes high, the RMSD takes large values, and when the temperature becomes low, the RMSD takes small values. By comparing Figs. \ref{1afo-integrate}(b) and \ref{1afo-integrate}(d), we see that there is a positive correlation between the temperature and the RMSD values. The fact that the RMSDs at high temperatures are large implies that our simulations did not get trapped in local-minimum potential-energy states.
These results confirm that the REM simulation was properly performed.

\begin{figure}[htb]
\centering
\includegraphics[width=.9\linewidth]{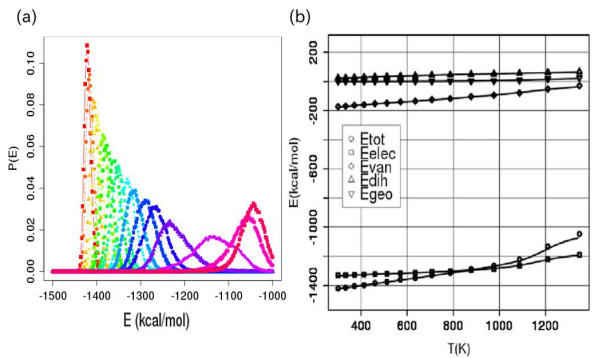}
\caption{\label{1afo-wham}(a) Canonical probability distributions of total potential energy at each temperature from the REM simulation of glycophorin A. The distributions correspond to the following temperatures (from left to right): 300, 333, 371, 413, 460,  512, 571, 635, 707, 787, 877, 976, 1087, 1210, 1347, and 1499 K. (b) The averages of the total potential energy Etot of glycophorin A and its component terms: electrostatic energy Eele, van der Waals Evdw, dihedral energy Edih, and  constraint energy Egeo as functions of temperature.}
\end{figure}

Table \ref{1afo accept} lists the acceptance ratios of replica exchange between
all pairs of nearest neighboring temperatures.
We find that the acceptance ratio is high enough ($>$ 0.1) in all
temperature pairs.
Fig. \ref{1afo-wham}(a) shows the canonical  probability distributions of the potential
energy obtained from the REM simulation at 16 temperatures.
We see that the distributions have enough overlaps between the neighboring
temperature pairs. This ensures that the number of replicas was sufficient.
In Fig. \ref{1afo-wham}(b), the average potential energy and its components, namely, the electrostatic energy $E_{elec}$, van der Waals energy $E_{vdw}$, torsion energy $E_{dih}$, and constraint energy $E_{geo}$, are shown as functions of temperature, which were calculated by eq. (\ref{wham}).
Because the helices are generally far apart from each other at high temperatures, the energy components, especially electrostatic energy and van der Waals energy, are higher at high temperatures.
At low temperatures, on the other hand, the side-chain packing among helices is expected.
We see that as the temperature becomes lower, $E_{vdw}$, $E_{dih}$, and $E_{elec}$ decrease almost linearly up to $\sim$ 1200 K, and as a result $E_{tot}$ is also almost linearly decreasing up to $\sim$ 1200 K. On the other hand, when the temperature becomes $<$ 1200 K, $E_{vdw}$ contributes more to the decrease of $E_{tot}$.
This is reasonable, because $E_{vdw}$ decreases as a result of side-chain packing and the stability of the conformation increases. 
Note that we used only transmembrane regions in the REM simulation. Transmembrane helices are generally considered to be hydrophobic, and helix-helix association is sometimes considered only by vdW packing (lock-and-key model). However, Fig. \ref{1afo-wham}(b) shows that $E_{elec}$ also changes much as a function of temperature. This implies that electrostatic effects also contribute to the formation of the native protein conformation.

%
\begin{table}[!tbp]
\caption{Acceptance ratios of replica exchange corresponding to pairs of
 neighboring temperatures from the REM simulation of glycophorin A.\label{1afo accept}} 
\begin{center}
\begin{tabular}{lclc}
\hline
\multicolumn{1}{c}{Pairs of $T$ }&\multicolumn{1}{c}{Acceptance ratio}&\multicolumn{1}{c}{Pairs of $T$ }&\multicolumn{1}{c}{Acceptance ratio}\tabularnewline
\hline
 300 $\longleftrightarrow$   333    &$0.43$&  707  $\longleftrightarrow$   787    &$0.41$\tabularnewline
 333  $\longleftrightarrow$   371    &$0.42$& 787  $\longleftrightarrow$   877    &$0.39$\tabularnewline
 371  $\longleftrightarrow$   413    &$0.41$& 877  $\longleftrightarrow$   976    &$0.39$\tabularnewline
 413  $\longleftrightarrow$   460    &$0.42$& 976  $\longleftrightarrow$  1087    &$0.30$\tabularnewline
 460  $\longleftrightarrow$   512    &$0.43$&1087  $\longleftrightarrow$  1210    &$0.14$\tabularnewline
 512  $\longleftrightarrow$   571    &$0.42$&1210  $\longleftrightarrow$  1347    &$0.20$\tabularnewline
 571  $\longleftrightarrow$   635    &$0.43$&1347  $\longleftrightarrow$  1499    &$0.40$\tabularnewline
 635  $\longleftrightarrow$   707    &$0.42$&&\tabularnewline
\hline
\end{tabular}
\end{center}
\end{table}

\subsubsection*{Principal component analysis}
\label{sec-3-1-2}

\begin{figure}[htb]
\centering
\includegraphics[width=0.4\textwidth]{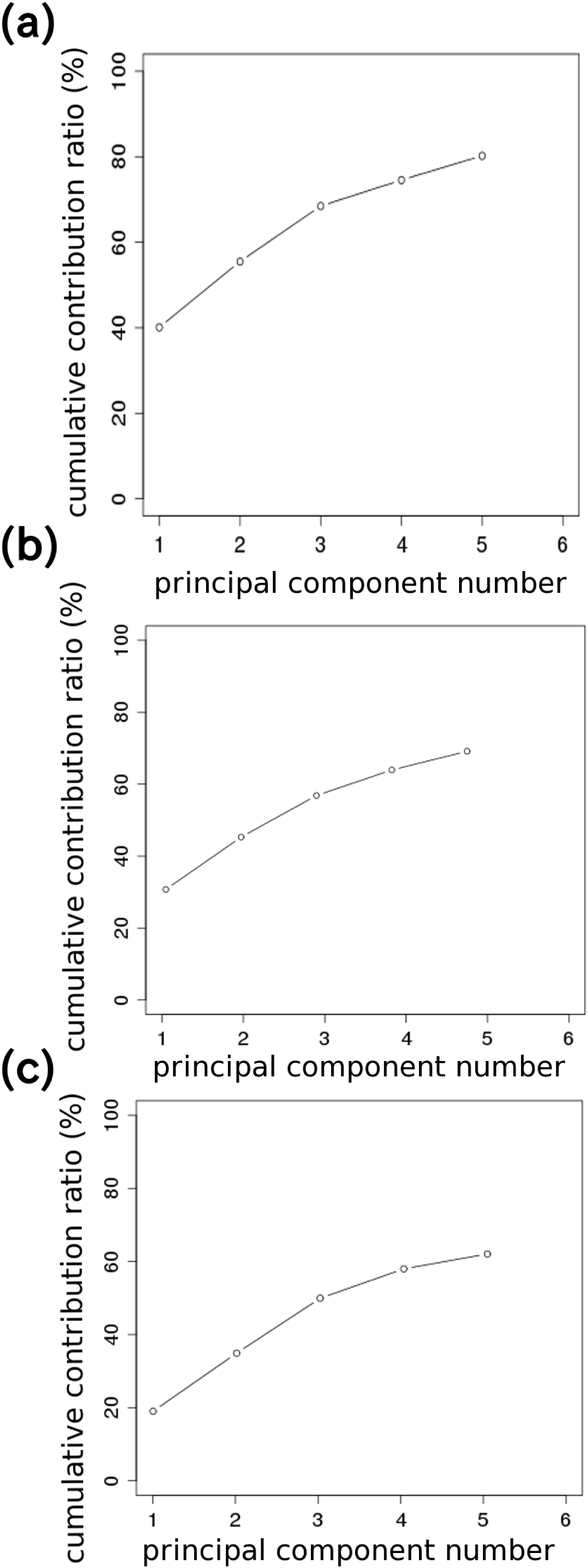}
\caption{\label{1afo-cumu}Cumulative contribution ratio of the first five eigenvalues in the principal component analysis from sampled structures of the REM simulation of glycophorin A at 300 K (a), 635 K (b), 1499 K (c).}
\end{figure}

\begin{figure}[htb]
\centering
\includegraphics[width=.9\linewidth]{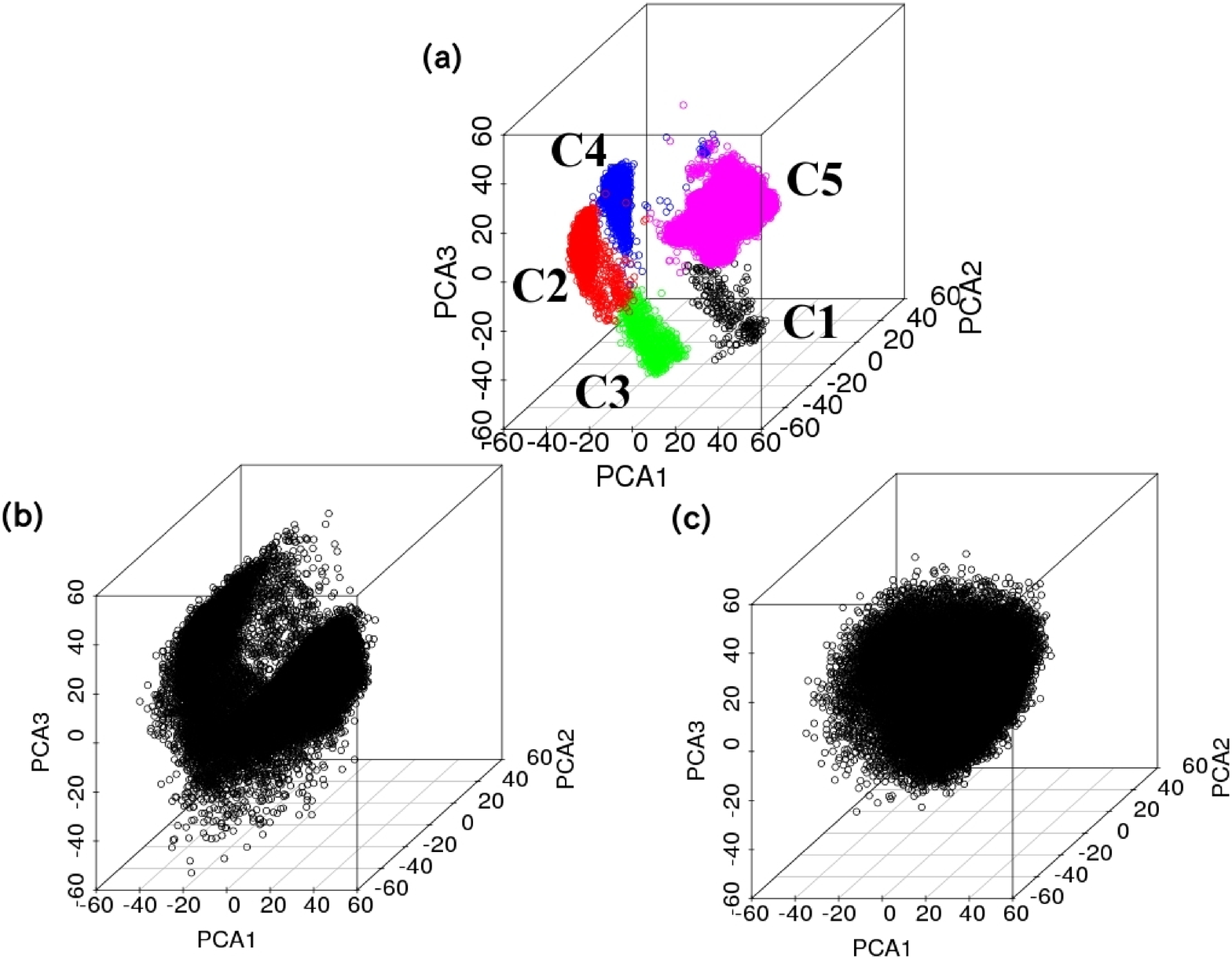}
\caption{\label{1afo-cluster3d300}Projection of sampled structures on the first, second, third principal axis from the REM simulation glycophorin A at temperature 300 K (a), 635 K (b), 1499 K (c). PCA1, PCA2, and PCA3 represent the principal axes 1, 2 and 3, respectively. Only structures in (a) are classified into clusters of similar structures and analyzed in detail. In panel (a), C1, C2, $\cdots$ , C5 stand for Cluster 1, 2, $\cdots$ , 5, respectively, and are highlighted by different colors.}
\end{figure}

We now classify the sampled structures into clusters of similar structures by the principal component analysis.
In Fig. \ref{1afo-cumu}, we show the percentage of the cumulative contribution ratio of the first five eigenvalues at the chosen temperature of 300 K (lowest), 707 K, and 1499 K (highest). We see from the ratio values in Fig. \ref{1afo-cumu} that as the temperature becomes higher, more principal component axes are needed to represent the fluctuations of the structures, as is expected. This is reasonable because as the temperature becomes higher, the fluctuations of the system become larger and the simulation samples a wider conformational space. In Fig. \ref{1afo-cumu}(a), we see that more than 60 \% of the total fluctuations at 300 K is expressed by the first three principal components. Although we can express the system more precisely as we use more principal axes, we here classify and analyze the sampled structures at the lowest temperature by the first three principal components. 
The fact that most of the amplitudes of fluctuations in this protein system is represented only by a small number of principal components is consistent with that protein folding dynamics can be expressed as the diffusion over a low-dimensional free energy surface as is elucidated in the energy landscape theory.
Fig. \ref{1afo-cumu}(c) shows that many principal component axes are needed to express the sampled structures properly at the highest temperature. The sampled structures are sometimes analyzed by other reaction coordinates such as native contact, RMSD, and radius of gyration. These are suitable as reaction coordinates in some cases but may not be appropriate in others. We do not know how many reaction coordinates we need for identifying important local-minimum free energy states in the free energy landscape. The principal component analysis is one of the methods that naturally provide us with the information as to how many reaction coordinates we need for such investigations.
In Fig. \ref{1afo-cluster3d300}, the projection of sampled structures from the REM simulation on the first, second, and third principal component axes (PCA) at the chosen three temperatures is shown.
In Fig. \ref{1afo-cluster3d300}(a), each cluster of structures is highlighted with different colors.
If we perform constant temperature simulations at the lowest temperature, the simulations will get trapped in one of the clusters in Fig \ref{1afo-cluster3d300}(a), depending on the initial conformations of the simulations. However, each replica of the replica-exchange simulation will not get trapped in one of the local-minimum free energy states, by going through high temperature regions. Every replica can climb over energy barriers in Fig. \ref{1afo-cluster3d300}(c) by temperature exchange during the simulation. This is the reason why we adopted the replica-exchange method.
At the lowest temperature, we classified sampled structures at the lowest temperature into five distinct clusters in Fig. \ref{1afo-cluster3d300}(a). They lie in the ranges  (--13 --- 16; 2 --- 32; --77 --- --19), (--49 --- --13; --34 --- --2; --34 --- 13), (--35 --- 8; --21 --- 16; --81 --- --30), (--53 --- --14; --7 --- 39; --30 --- 89), and (--20 --- 28; --37 --- 37; --24 --- 81), which we refer to as Cluster 1, Cluster 2, Cluster 3, Cluster 4, and Cluster 5, respectively.

\subsubsection*{Average quantities of clusters}
\label{sec-3-1-3}

Table \ref{1afo allcluster} lists average quantities of five clusters of similar structures.
These structures were extracted from the trajectories at a fixed interval.
The rows of Cluster 1, Cluster 2,$\cdots$ , and Cluster 5 represent various average values for the structures that belong to each cluster.
We see that in the value of the "Str" column in Table \ref{1afo allcluster}, Cluster 5 has the most number of structures.
Hence, Cluster 5 is the global-minimum free energy state in this simulation.
Others are considered to be local-minimum free energy states.
As for the RMSD values, Cluster 5 has the value 2.67 \AA{}, and it is the lowest value among the five clusters. Therefore, Cluster 5 corresponds to the global-minimum free energy state and it is also closest to the native structure.

\begin{table}[!tbp]
\caption{Various average quantities of glycophorin A for each cluster at the temperature of 300 K.\label{1afo allcluster}} 
\begin{center}
\begin{tabular}{lrrrrrrr}
\hline
\multicolumn{1}{l}{}&\multicolumn{1}{c}{Str}&\multicolumn{1}{c}{Tote}&\multicolumn{1}{c}{Elec}&\multicolumn{1}{c}{Vdw}&\multicolumn{1}{c}{Dih}&\multicolumn{1}{c}{Geo}&\multicolumn{1}{c}{RMSD}\tabularnewline
\hline
   Cluster 1&   $  325$&   $-1414$&   $-1332$&   $-163$&   $23.3$&   $0.982$&   $7.04$\tabularnewline
   Cluster 2&   $ 2583$&   $-1419$&   $-1329$&   $-173$&   $24.2$&   $0.851$&   $6.75$\tabularnewline
   Cluster 3&   $ 1349$&   $-1422$&   $-1329$&   $-174$&   $22.5$&   $0.697$&   $6.95$\tabularnewline
   Cluster 4&   $ 5433$&   $-1422$&   $-1328$&   $-177$&   $24.9$&   $0.660$&   $6.67$\tabularnewline
   Cluster 5&   $50309$&   $-1421$&   $-1330$&   $-173$&   $23.0$&   $0.670$&   $2.67$\tabularnewline
\hline
\end{tabular}
\end{center}
The following abbreviations are used: Str: the number of structures,
Tote: total potential energy, Elec: electrostatic energy, Vdw: van
der Waals energy, Dih: dihedral energy, Geo: constraint energy (all in
kcal/mol), RMSD: root-mean-square deviation of all C$^\alpha$ atoms (in \AA). 
\end{table}

\begin{figure}[htb]
\centering
\includegraphics[width=.9\linewidth]{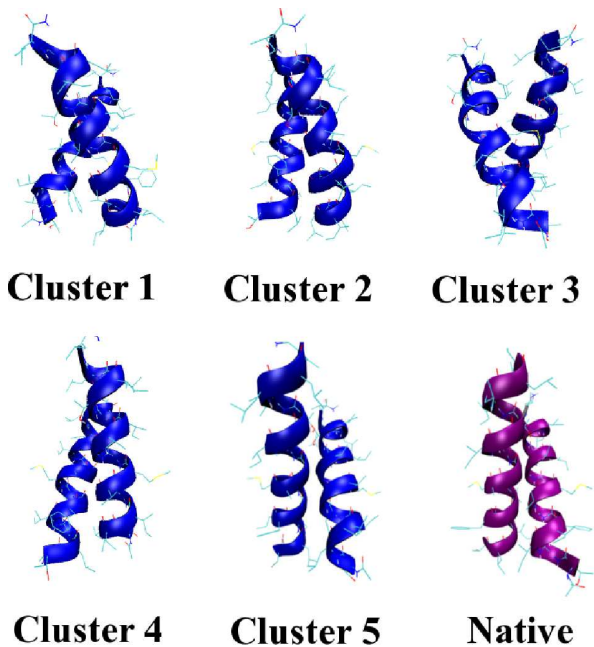}
\caption{\label{1afo-minimumenestr}(Color online) Typical structures of glycophorin A in each cluster selected by  free energy local minimum state. The purple structure is native structure. The RMSD from the native conformation with respect to all the C$^{\alpha}$ atoms is 6.80, 6.65, 7.15, 6.52, and 2.25 \AA{} for Cluster 1, Cluster 2, Cluster 3, Cluster 4, and Cluster 5, respectively.}
\end{figure}

We next examine the typical local-minimum free energy state structures in each cluster. 
The representative structures were selected in the highest density regions within the clusters.
In Fig. \ref{1afo-minimumenestr}, the representative structure of each cluster and the solution NMR structure (PDB code:1AFO) are shown. We confirm that the structure of Cluster 5 is the closest to the experimental  one.
Note that each helix in all these structures has a similar structure, which is close to an ideal helix structure. This means that glycophorin A has only ideal helix structures as local-minimum free energy states in this simulation, although we allowed helix to be distorted or bent during the simulation.

\subsection*{Phospholamban}
\label{sec-3-2}
\subsubsection*{Time series of variety quantities}
\label{sec-3-2-1}

\begin{figure}[htb]
\centering
\includegraphics[width=.9\linewidth]{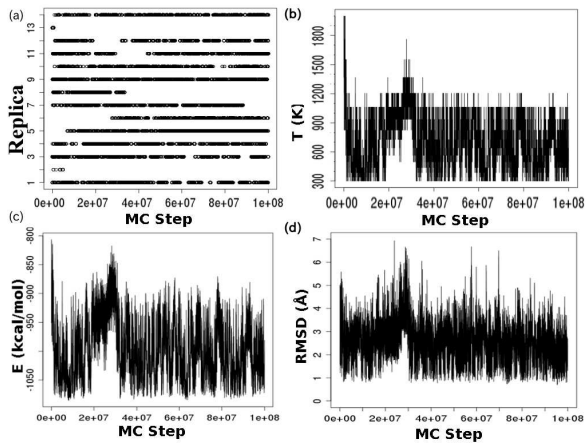}
\caption{\label{1fjk-integrate}Time series of various quantities for the REM simulation of phospholamban. (a) Time series of replica index at temperature 300 K. (b) Time series of temperature change for Replica 12. (c) Time series of total potential energy change for Replica 12. (d) Time series of the RMS deviation (in \AA{}) of all the C$^{\alpha}$ atoms from the PDB structures for Replica 12.}
\end{figure}

We next examine how the replica-exchange simulation performed for phospholamban. 
Fig. \ref{1fjk-integrate}(a) shows the time series of the replica index at the lowest temperature of 300 K.
We see that many replicas experience the minimum temperature many times
during the REM simulation, and a random walk in the replica space was realized.
The complementary picture is the temperature exchange for each replica.
Fig. \ref{1fjk-integrate}(b) shows the results for one of the replicas (Replica 12). We see that 
Replica 12 reached various temperatures  during the REM simulation and the random walk in the temperature space between the lowest and highest temperatures was also realized. Other replicas behaved similarly.
Fig. \ref{1fjk-integrate}(c) shows the corresponding time series of  the total potential energy. 
We next examine how widely the conformational space was sampled during the REM simulation.
We plot the time series of RMSD of all the C$^{\alpha}$ atoms from the experimental structure (PDB code: 1FJK) for Replica 12 in Fig. \ref{1fjk-integrate}(d).
When the temperature becomes high, the RMSD takes large values, and when the temperature becomes low, the RMSD takes small values. By comparing Figs. \ref{1fjk-integrate}(b), \ref{1fjk-integrate}(c), and \ref{1fjk-integrate}(d), we see that there is a strong correlation among the temperature, total potential energy and RMSD values. The fact that RMSD at high temperatures is large implies that our simulations did not get trapped in local-minimum potential-energy states.
These results confirm that the REM simulation was properly performed.

\begin{figure}[htb]
\centering
\includegraphics[width=.9\linewidth]{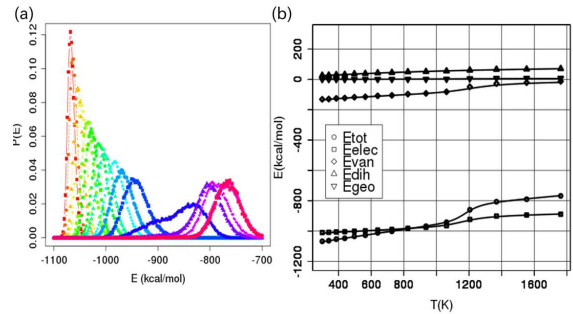}
\caption{\label{1fjk-wham}(a) Canonical probability distributions of total potential energy at each temperature from the REM simulation of phospholamban. The distributions correspond to the following temperatures (from left to right): 300, 340, 386, 438, 497, 564, 640, 727, 825, 936, 1062, 1205, 1368, 1553, 1762, and 2000 K. (b) The averages of the total potential energy Etot of glycophorin A and its component terms: electrostatic energy Eele, van der Waals Evdw, dihedral energy Edih, and  constraint energy Egeo as functions of temperature.}
\end{figure}

Table \ref{1fjk accept} lists the acceptance ratios of replica exchange between
all pairs of nearest neighboring temperatures.
We find that almost all acceptance ratios are high enough 
($>$ 0.1) 
in temperature pairs.
Fig. \ref{1fjk-wham}(a) shows canonical  probability distributions of the potential
energy obtained from the REM simulation at 16 temperatures.
We see that the distributions have enough overlaps between the neighboring
temperature pairs. This ensures that the number of replicas was sufficient.
We also see that the fourth highest distribution is broader than other distributions, and it  suggests that the phase transition occurs around this temperature (1205 K). 
In Fig. \ref{1fjk-wham}(b), the average potential energy and its components (the electrostatic energy $E_{elec}$, van der Waals energy $E_{vdw}$, torsion energy $E_{dih}$, and constraint energy $E_{geo}$) are shown as functions of temperature, which were calculated by eq. (\ref{wham}).
Because the helix is distorted at high temperatures, the energy components, especially electrostatic energy and van der Waals energy, are higher at high temperatures.
At low temperatures, on the other hand, we observe the formation of helix structures.
We see that the change of energy components behaves similarly of glycophorin A.
However, the drastic change at about 1200 K of phospholamban may be affected by the low acceptance ratio of the pair between 1062 K and 1205 K.
All the energy components contribute to the formation of the native helix conformation.

%
\begin{table}[!tbp]
\caption{Acceptance ratios of replica exchange corresponding to pairs of
 neighboring temperatures from the REM simulation of phospholamban.\label{1fjk accept}} 
\begin{center}
\begin{tabular}{lclcr}
\hline
\multicolumn{1}{c}{Pairs of $T$ }&\multicolumn{1}{c}{Acceptance ratio}&\multicolumn{1}{c}{Pairs of $T$ }&\multicolumn{1}{c}{Acceptance ratio}\tabularnewline
\hline
 300  $\longleftrightarrow$   340    &$0.42$& 825  $\longleftrightarrow$   936    &$0.41$\tabularnewline 
 340  $\longleftrightarrow$   386    &$0.41$& 936  $\longleftrightarrow$  1062    &$0.37$\tabularnewline
 386  $\longleftrightarrow$   438    &$0.41$&1062  $\longleftrightarrow$  1205    &$0.08$\tabularnewline
 438  $\longleftrightarrow$   497    &$0.42$&1205  $\longleftrightarrow$  1368    &$0.36$\tabularnewline
 497  $\longleftrightarrow$   564    &$0.42$&1368  $\longleftrightarrow$  1553    &$0.39$\tabularnewline
 564  $\longleftrightarrow$   640    &$0.43$&1553  $\longleftrightarrow$  1762    &$0.32$\tabularnewline
 640  $\longleftrightarrow$   727    &$0.42$&1762  $\longleftrightarrow$  2000    &$0.11$\tabularnewline
 727  $\longleftrightarrow$   825    &$0.42$&&\tabularnewline
\hline
\end{tabular}
\end{center}
\end{table}

\subsubsection*{Principal component analysis}
\label{sec-3-2-2}
\begin{figure}[htb]
\centering
\includegraphics[width=0.4\textwidth]{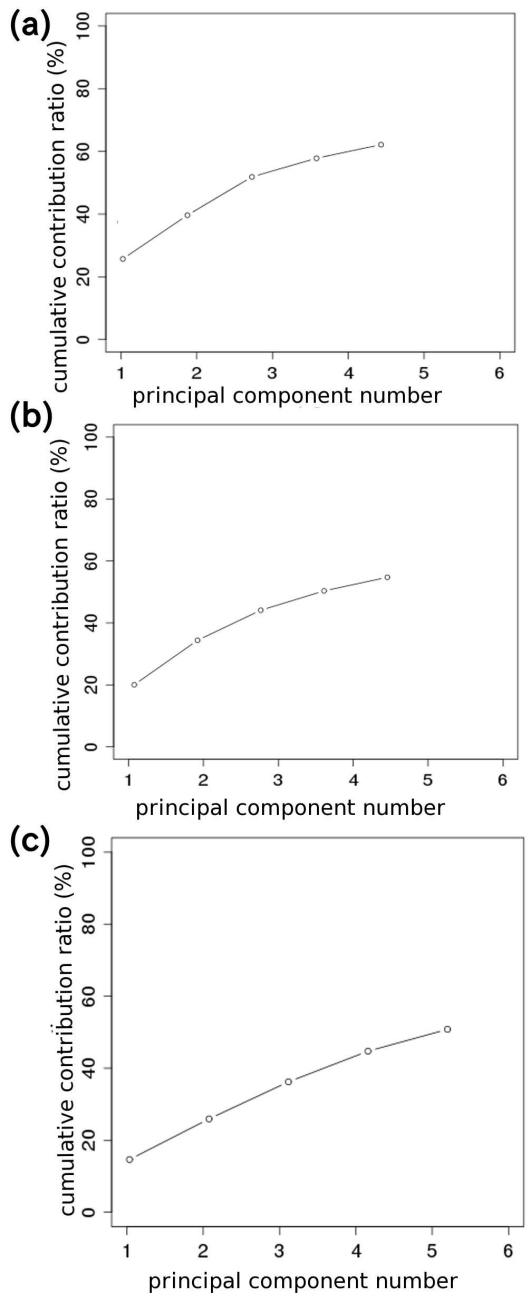}
\caption{\label{1fjk-cumu}Cumulative contribution ratio of the first five eigenvalues in the principal component analysis from sampled structures of the REM simulation of phospholamban at 300 K (a), 936 K (b), 2000 K (c).}
\end{figure}

\begin{figure}[htb]
\centering
\includegraphics[width=.9\linewidth]{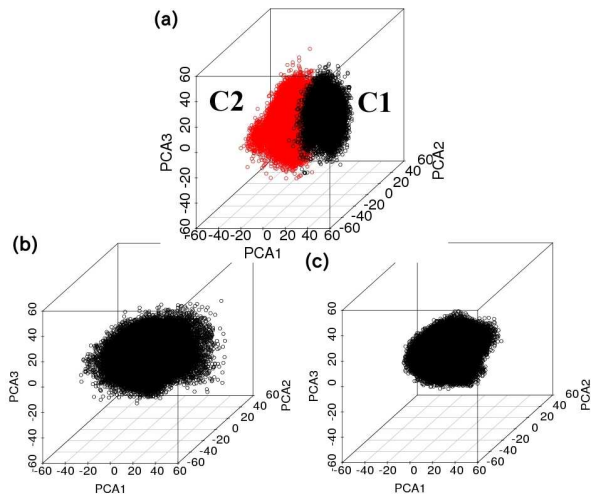}
\caption{\label{1fjk-cluster3d300}Projection of sampled structures on the first, second, third principal axis from the REM simulation phospholamban at temperature 300 K (a), 936 K (b), 2000 K (c). PCA1, PCA2, and PCA3 represent the principal axes 1, 2 and 3, respectively. Only structures in (a) are classified into clusters of similar structures and analyzed in detail. In panel (a), C1 and C2 stand for Cluster 1 and Cluster 2, respectively, and are highlighted by different colors.}
\end{figure}

We now classify the sampled structures into clusters of similar structures by the principal component analysis again.
In Fig. \ref{1fjk-cumu}, we show the percentage of the cumulative contribution ratio of the first five eigenvalues at the chosen temperatures of 300 K (lowest), 936 K, and 2000 K (highest). 
We see from the ratio values in Fig. \ref{1fjk-cumu} that as the temperature becomes higher, more principal component axes are needed to represent the fluctuations of the structures, as it is expected.
In Fig. \ref{1fjk-cumu}, we see that more than 50 \% of the total fluctuations at the lowest temperature is expressed by the first three principal components. The less ratio compared to that of glycophorin A may have resulted from the fact that phospholamban had many helix structures including distorted ones, whereas glycophorin A had mostly ideal helix structures.
Fig. \ref{1fjk-cumu}(c) shows that many principal component axes are needed to express the sampled structures properly at the highest temperature. 
In Fig \ref{1fjk-cluster3d300}, the projection of sampled structures from the REM simulation on the first, second, and third principal component axes (PCA) at the chosen three temperatures is shown.
In Fig. \ref{1fjk-cluster3d300}(a), each cluster of structures is highlighted with different colors.
Every replica can climb over energy barriers in Fig. \ref{1fjk-cluster3d300}(c) by temperature exchange during the REM simulation. 
Compared with the results of glycophorin A, the distribution of sampled structures projected by three principal components is simple, and one principal component axis already distinguishes the clusters. This may result from the fact that we have only a helix in the system without helix-helix interaction.
 Our simulation can not sample random coil structures and the conformational space is restricted into a narrow space. 
At the lowest temperature, we classified sampled structures at the lowest temperature into two distinct clusters in Fig. \ref{1fjk-cluster3d300}(a). They lie in the ranges  (5 --- 50; --35  --- 32; --36 --- 39) and (--36 --- 13; --49 --- 40; --42 --- 50), which we refer to as Cluster 1 and Cluster 2, respectively.

\subsubsection*{Average quantities of clusters}
\label{sec-3-2-3}

\begin{table}[!tbp]
\caption{Various average quantities of phospholamban for each cluster at
 the temperature of 300 K. \label{1fjk allcluster}} 
\begin{center}
\scalebox{0.95}[1]{ 
\begin{tabular}{lrrrrrrr}
\hline
\multicolumn{1}{l}{}&\multicolumn{1}{c}{Str}&\multicolumn{1}{c}{Tote}&\multicolumn{1}{c}{Elec}&\multicolumn{1}{c}{Vdw}&\multicolumn{1}{c}{Dih}&\multicolumn{1}{c}{Geo}&\multicolumn{1}{c}{RMSD}\tabularnewline
\hline
   Cluster 1&   $20519$&   $-1064$&   $-1011$&   $-126$&   $24.8$&   $0.23$&   $2.06$\tabularnewline
   Cluster 2&   $79480$&   $-1067$&   $-1009$&   $-132$&   $25.8$&   $0.26$&   $2.93$\tabularnewline
\hline
\end{tabular}
}
\end{center}
The following abbreviations are used: Str: the number of structures,
Tote: total potential energy, Elec: electrostatic energy, Vdw: van
der Waals energy, Dih: dihedral energy, Geo: constraint energy (all in
kcal/mol), RMSD: root-mean-square deviation of all C$^\alpha$ atoms (in \AA). 
\end{table}

\begin{figure}[htb]
\centering
\includegraphics[width=.9\linewidth]{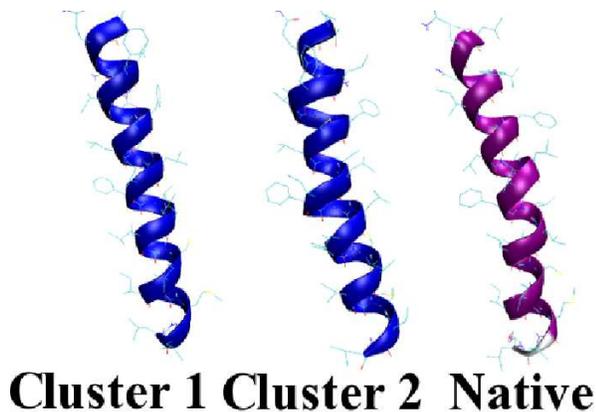}
\caption{\label{1fjk-minimumenestr}(Color online) Typical structures of phospholamban in each cluster selected by  free energy local minimum state. The purple structure is native structure. The RMSD from the native conformation with respect to the backbone atoms is 1.27 and 2.89 \AA{} for Cluster 1 and Cluster 2, respectively.}
\end{figure}

Table \ref{1fjk allcluster} lists average quantities of two clusters of similar structures.
The rows of Cluster 1 and Cluster 2 represent various average values for the structures that belong to each cluster.
We see that RMSD is as small as 2.06 \AA{} for Cluster 1, while
it is 2.93 \AA{} for Cluster 2. Hence, Cluster 1 has very similar
structures to the native one. However, it is not the
global-minimum free energy state but a local-minimum one,
comparing the number of conformations (Str entries in
Table \ref{1fjk allcluster}) in both clusters.

In Fig. \ref{1fjk-minimumenestr}, representative structures of each cluster
in Table \ref{1fjk allcluster} and the structure obtained by solution NMR
experiments (PDB code: 1FJK) are shown. We confirm that
Cluster 1 is very similar to the native structure. It
is bent at the same position and in the same direction,
although the amount of bent is not as much as the native
one. Cluster 2 is also bent at the same position and
about the same amount as the native one, but it has a
bend in the opposite direction. Hence, the present
simulation can predict the position of bend, but it
gives both directons of bend as local-minimum free
energy states and Cluster 2 as the global-minimum one.
The present system is a helix monomar, and without
interactions with other helices, it seems very difficult
to decide the direction of distorsions within the
approximation of the present method.
We remark that a preliminary REM simulation of
bacteriorhodopsin with seven helices predicts correct
directions of helix bending (manuscript in preparation).

\section*{Conclusions}
\label{sec-4}
In this article, we introduced deformations of helix structures to the replica-exchange Monte Carlo simulation for membrane protein structure predictions.
The membrane bilayer environment was approximated by restraining the conformational space in virtual membrane region. The sampled helix structures were limited so that helix structures by introducing the restraints on the backbone $\phi$ and $\psi$ angles are not completely destroyed.
In order to check the effectiveness of the method, we first applied it to the prediction of a dimer membrane protein, glycophorin A. 
We successfully reproduced the native-like structure as the global-minimum free energy state.
We next applied the method to phospholamban, which has one distorted transmembrane helix structure in the PDB structure. 
The results implied that a native-like structure was obtained as a local-minimum free energy state. Two local-minimum free energy states were found with the same bend position as the native one, but the global-minimum free energy state had an opposite direction of helix bend.
Therefore, our results seem to imply  that the location of bends of helix structures in transmembrane helices are determined by their amino-acid sequence, but the direction and amount of distortion of helices are dependent on the interactions with surrounding lipid molecules, which we represented only implicitly.
Our next targets will be more complicated membrane proteins with multiple transmembrane helices such as G protein coupled receptors. Our preliminary results for bacteriorhodopsin show that native-like structures with the correctly bent helices can be predicted by our method.

\section*{Acknowledgements}
\label{sec-5}
Some of the computations were performed on the supercomputers at the
Institute for Molecular Science, at the Supercomputer Center, Institute
for Solid State Physics, University of Tokyo, and Center for
Computational Sciences, University of Tsukuba.
This work was supported, in part, Grants-in-Aid for Scientific Research (A) (No. 25247071),
for Scientific Research on Innovative Areas (\lq\lq Dynamical Ordering \& Integrated
Functions\rq\rq ), Program for Leading Graduate Schools \lq\lq Integrative Graduate Education and Research in Green Natural Sciences\rq\rq, and for the Computational Materials Science Initiative, and for High Performance Computing Infrastructure from the Ministry of
Education, Culture, Sports, Science and Technology (MEXT), Japan.


\end{document}